\documentclass[12pt]{article}
\usepackage[utf8]{inputenc}
\usepackage{amsfonts}
\usepackage{amssymb}
\usepackage{amsmath}
\usepackage{amstext}
\usepackage{amsthm}
\usepackage{array}
\usepackage{chngpage}
\usepackage{graphicx}
\usepackage[margin=1.25in]{geometry}
\usepackage{mathptmx}
\usepackage[colorlinks=false,pdfborder={0 0 0}]{hyperref}
\usepackage[longnamesfirst]{natbib}
\usepackage{setspace}
\usepackage{multirow}
\usepackage{booktabs}
\usepackage{threeparttable}
\usepackage{xcolor}
\usepackage{lscape}
\usepackage[title]{appendix}
\providecommand{\U}[1]{\protect\rule{.1in}{.1in}}
\theoremstyle{plain}
\newtheorem{theorem}{Theorem}

\newtheorem{proposition}[theorem]{Proposition}
\theoremstyle{definition}

\newcommand{\pr}{\mathrm{P}}
\newcommand{\e}{\mathrm{E}}

\newcommand{\bge}{\begin{equation}}
	\newcommand{\ene}{\end{equation}}

\def\monthname{\ifcase\month\or January\or February\or March\or April\or May\or June\or July\or August\or September\or October\or November\or December\fi}

\usepackage{titlesec}

\titleformat{\section}
{\normalfont\normalsize\bfseries\centering}
{\thesection}{1em}{}

\titleformat{\subsection}
{\normalfont\normalsize\bfseries\centering}
{\thesubsection}{1em}{}

\titleformat{\subsubsection}
{\normalfont\normalsize\bfseries\centering}
{\thesubsubsection}{1em}{}

\titlespacing*{\section}{0pt}{0.8\baselineskip}{0.45\baselineskip}
\titlespacing*{\subsection}{0pt}{0.75\baselineskip}{0.4\baselineskip}
\titlespacing*{\subsubsection}{0pt}{0.7\baselineskip}{0.35\baselineskip}

\expandafter\def\expandafter\normalsize\expandafter{%
	\normalsize
	\setlength\abovedisplayskip{8pt}
	\setlength\belowdisplayskip{8pt}
	\setlength\abovedisplayshortskip{2pt}
	\setlength\abovedisplayshortskip{2pt}
}

\begin{document}
	
	\title{\textsc{\Large{Covariate Balancing and the Equivalence of Weighting and Doubly Robust Estimators of Average Treatment Effects}}\footnote{This version:~September 19, 2025. This paper was presented at Brandeis University, Ege University, the \mbox{2025 Annual} Conference of the International Association for Applied Econometrics, the 2025 Workshop in Econometrics at TU Dortmund University, the 2024 LMU Munich - Todai Econometrics Workshop, the 2024 Econometric Society European Meetings, the 2024 BSE Summer Forum Workshop on Microeconometrics and Policy Evaluation, the 2023 BU/BC Greenline Econometrics Workshop, and the 2023 Annual Meeting of the Southern Economic Association. We thank the audiences at those seminars and conferences, as well as Bernie Black, Brant Callaway, Bruno Ferman, Bryan Graham, Pedro Sant'Anna, Rahul Singh, Max Tabord-Meehan, Joachim Winter, Yinchu Zhu, and Jos\'e Zubizarreta, for helpful conversations and comments. We also thank Lennart Ohly for excellent research assistance.}}
	
	\author{\textsc{Tymon S{\l}oczy\'nski}\thanks{Department of Economics, Brandeis University, \texttt{tslocz@brandeis.edu}}
		\and
		\textsc{S. Derya Uysal}\thanks{Department of Economics, LMU Munich, \texttt{derya.uysal@econ.lmu.de}}
		\and
		\textsc{Jeffrey M. Wooldridge}\thanks{Department of Economics, Michigan State University, \texttt{wooldri1@msu.edu}}
	}
	
	\date{}
	
	\maketitle
	
	\begin{abstract}
		\noindent
		How should researchers adjust for covariates? We show that if the propensity score is estimated using a specific covariate balancing approach, inverse probability weighting (IPW), augmented inverse probability weighting (AIPW), and inverse probability weighted regression adjustment (IPWRA) estimators are numerically equivalent for the average treatment effect (ATE), and likewise for the average treatment effect on the treated (ATT)\@. The resulting weights are inherently normalized, making normalized and unnormalized IPW and AIPW identical. We discuss implications for instrumental variables and difference-in-differences estimators and illustrate with two applications how these numerical equivalences simplify analysis and interpretation.
	\end{abstract}
	
	\begin{small}
		 \noindent
		
	\end{small}
	
	\bigskip\bigskip\bigskip
	\bigskip\bigskip\bigskip
	\bigskip\bigskip\bigskip
	
	\thispagestyle{empty}
	
	\pagebreak
	
	\onehalfspacing
	\setcounter{page}{2}

	\section{Introduction}
	\label{sec:intro}
	
	Covariate adjustment is central to causal inference, yet the choice of method remains contested. Much recent research has highlighted the shortcomings of a number of well-established estimation methods in reproducing suitable averages of heterogeneous treatment effects. A key lesson from this literature is that additive linear models may often fail to properly adjust for covariates when those covariates are relevant for identification. This concern arises not only under unconfoundedness \citep[e.g.,][]{Sloczynski2022,GPHK2024,Chen2025}, but also in instrumental variables \citep[e.g.,][]{Sloczynski2024,BBMT2025} and difference-in-differences settings \citep[e.g.,][]{CC2024}.
	
	When considering alternatives to standard methods, researchers face a wide range of options, including regression adjustment, matching, weighting, and doubly robust estimators, as well as related approaches based on machine learning. Such variety is not necessarily desirable: the existence of a common standard facilitates comparability across studies, while additional researcher degrees of freedom invite specification searching \citep{SNS2011,Vivalt2019,FPP2020}.
	
	In this paper, we build on the fact that many of these alternative methods require first-step estimation of the propensity score. We assume that a researcher would be willing to commit to estimating the propensity score using a particular method of moments approach with desirable properties, namely the inverse probability tilting (IPT) estimator of \cite{EGP2008} and \cite{GPE2012,GPE2016}. (To be clear, a researcher using IPT still needs to choose a model for the propensity score, perhaps logit or probit, but would then estimate this model using the method of moments instead of maximum likelihood.) Our main contribution is to demonstrate that commitment to using IPT substantially reduces the choice set (i.e., the number of alternative estimators) available to researchers. Specifically, we show that using the IPT moment conditions to estimate the propensity score leads to numerical equivalence between members of several classes of estimators of average treatment effects: inverse probability weighting (IPW), augmented inverse probability weighting (AIPW), and inverse probability weighted regression adjustment (IPWRA), with the latter two classes using linear models for potential outcome means. Our equivalence results are very general, as they are valid for any propensity score model having the index form (e.g., logit or probit). In addition, they apply to both normalized and unnormalized versions of IPW and AIPW, as well as to estimators of both the average treatment effect (ATE) and the average treatment effect on the treated (ATT)\@.
	
	Our equivalence results are meaningful insofar as IPT is appealing in its own right. Indeed, a major reason for this appeal is that the moment conditions underlying IPT impose a desirable property known as ``exact balancing.'' Specifically, IPT estimates the parameters of the propensity score model such that, after reweighting, the mean covariate values are identical across key groups: between the (weighted) treatment group, the (weighted) comparison group, and the (unweighted) full sample when estimating the ATE, and between the (unweighted) treatment group and the (weighted) comparison group when estimating the ATT\@. In this sense, IPT is a prototypical ``covariate balancing'' estimator, ensuring that causal comparisons are made only between groups with identical (mean) characteristics. As shown by \cite{EGP2008} and \cite{GPE2012,GPE2016}, IPT estimators of the ATE and ATT also enjoy local efficiency and double robustness properties. That is, if both the propensity score model chosen by the researcher (e.g., logit or probit) and the linear specification of potential outcome means are correct, the estimators are semiparametrically efficient; if only one is correctly specified, the estimators remain consistent. As we review below, the IPT estimator of the ATT is also identical to the subsequent proposals of \cite{Hainmueller2012} and \cite{IR2014}.\footnote{Specifically, IPT coincides with \citeauthor{Hainmueller2012}'s (\citeyear{Hainmueller2012}) entropy balancing estimator when the propensity score is estimated with the logit model.}
	
	We also translate our equivalence results to instrumental variables and difference-in-differences settings. In both contexts, weighting and doubly robust estimators have played a prominent role in the recent literature, and our results again simplify the set of alternative estimators available to empirical researchers. In particular, our results imply that the doubly robust estimator proposed by \cite{SAZ2020}, which uses IPT-estimated propensity scores, is numerically equivalent to the simple IPW estimator of \cite{Abadie2005} with IPT weights. It follows that in implementations of \cite{SAZ2020} it is redundant to estimate the model for the untreated potential change in outcomes.
	
	We illustrate our findings with two empirical applications. In the first application, we revisit the study of the causal effects of cash transfers on longevity in \cite{AEFLM2016}. In line with an earlier replication in \cite{Sloczynski2022}, we conclude that there is insufficient evidence to reject the null hypothesis of zero average effects. Our numerical equivalences simplify analysis and interpretation, and none of the IPT estimates is significantly different from zero despite usually having smaller standard errors than the corresponding estimates based on maximum likelihood. In the second application, we replicate \citeauthor{SAZ2020}'s (\citeyear{SAZ2020}) analysis of the NSW--CPS data, which was previously analyzed by \cite{LaLonde1986} and many others. We show that many of the estimates reported by \cite{SAZ2020} would have been identical to their preferred estimates had they used IPT to estimate the propensity score in all cases.
	
	We also supplement this paper with a companion Stata package, \texttt{teffects2}, available at the Statistical Software Components (SSC) Archive. Our package implements IPW, AIPW, and IPWRA estimators of the ATE and ATT under unconfoundedness, with several approaches to estimate the weights, including IPT\@. The package can also be used to estimate the ATT in difference-in-differences settings after a suitable transformation of the outcome variable. This provides a novel implementation of the doubly robust difference-in-differences (DRDID) estimator proposed by \cite{SAZ2020}.

	\subsection*{Literature Review}
	
	This paper builds on a body of research on estimating average treatment effects under the assumption of unconfoundedness. We focus on three classes of estimators: inverse probability weighting (IPW), as in \cite{HIR2003}, augmented inverse probability weighting (AIPW), as in \cite{RRZ1994}, and inverse probability weighted regression adjustment (IPWRA), as in \cite{Wooldridge2007} and \cite{SW2018}. These classes of estimators, as well as several others, were surveyed by \cite{IW2009}, \cite{AC2018}, and \cite{Uysal2024}.
	
	Each of the classes of estimators we consider requires a first-step estimation of the propensity score. This is typically done using maximum likelihood estimation (MLE) of a standard binary response model for treatment assignment (e.g., logit or probit). However, this estimation approach does not guarantee that any desirable balancing properties are satisfied in finite samples. In contrast, various ``covariate balancing'' estimators of the propensity score are explicitly constructed with these properties in mind; they are also tailored to the specific parameter of interest (e.g., ATE or ATT) to improve the statistical properties of the corresponding treatment effect estimator.
	
	Following the early work on inverse probability tilting (IPT) by \cite{EGP2008} and \cite{GPE2012,GPE2016}, many papers have proposed alternative covariate balancing procedures for estimating average treatment effects. \cite{Hainmueller2012} suggested estimating the inverse probability weights directly---subject to balancing and normalizing constraints---rather than estimating the propensity score first and then inverting it to obtain the weights. Known as ``entropy balancing,'' this procedure was designed to estimate the ATT and was later shown to be identical to IPT when the latter uses the logit model \citep{ZP2017,Tan2020}. \cite{IR2014} proposed using different moment conditions than those in \cite{EGP2008} and \cite{GPE2012} when estimating the ATE; however, the resulting estimator lacks some desirable theoretical properties of IPT, such as double robustness. On the other hand, \citeauthor{IR2014}'s (\citeyear{IR2014}) moment conditions for estimating the ATT are the same as in IPT, which implies that the resulting estimator is also equivalent to IPT (as well as to entropy balancing when the logit model is used). \cite{Zubizarreta2015} relaxed the exact balancing requirements of earlier methods and proposed estimating weights that minimize variance subject to approximate balancing constraints. \cite{Zhao2019} unified and generalized much of the earlier work by introducing a covariate balancing framework based on optimizing loss functions tailored to a given estimand. \cite{SASX2022} proposed estimating the propensity score by maximizing balance across the entire covariate distribution rather than in selected functions of the covariates.
	
	Most of the early work on covariate balancing focused on addressing missing data problems and estimating average treatment effects under unconfoundedness. However, recent research has also applied similar ideas to estimating various parameters of interest in difference-in-differences \citep{SAZ2020,CSA2021} and instrumental variables settings \citep{Heiler2022,SASX2022,SS2024,SUW2025}.
	
	This paper is also related to the important work of \cite{RSLGR2007}, \cite{Kline2011}, \cite{CZ2023}, and \cite{BSDFO2025} demonstrating numerical equivalences between regression adjustment and weighting estimators of average treatment effects, under the constraint that both the potential outcome means and the weights are linear in covariates. While the weights may generally be approximated as a linear function of a high-dimensional dictionary, as in \cite{CNS2022} and \cite{BSDFO2025}, the corresponding parametric restriction is unlikely to be plausible in low-dimensional settings, especially since it is equivalent to assuming an inverse linear model for the propensity score. In such settings, some of the estimated weights are likely to be negative, which invalidates the sample boundedness property of the resulting estimator \citep[cf.][]{RSLGR2007}.
	
	In this paper, we extend and generalize this earlier work by demonstrating that the numerical equivalences between the IPW, AIPW, and IPWRA estimators are driven by the IPT moment conditions rather than parametric restrictions on the propensity score or the weights. Unlike our paper, the results in \cite{RSLGR2007}, \cite{Kline2011}, \cite{CZ2023}, and \cite{BSDFO2025} are specific to the inverse linear model for the propensity score. (Most of these results are also limited to IPW\@.) Under this strong parametric restriction, which our paper does not make, IPW, AIPW, and IPWRA with IPT moment conditions are also numerically equivalent to (linear) regression adjustment.

	\subsection*{Plan of the Paper}
	
	We organize the paper as follows. In Section \ref{sec:balancing}, we review the estimation problems solved by the IPT weights for the ATE and the ATT and discuss some simple implications. In Section \ref{sec:equivalence}, we derive the equivalences among various IPT-based estimators of the ATE and then the IPT-based estimators of the ATT\@. We emphasize that since we are establishing numerical equivalences, we do not need to, and do not, state the assumptions under which the estimators are consistent. These have been covered elsewhere and are well known. In Section \ref{sec:implications}, we discuss the consequences that the algebraic equivalence results have for estimating local average treatment effects with instrumental variables and heterogeneous treatment effects in difference-in-differences settings. In Section \ref{sec:applications}, we discuss our empirical applications. In Section \ref{sec:conclusion}, we conclude. Our proofs are provided in the Appendix. In the Supplemental Appendix, we briefly discuss implementation of IPT in R and Stata.

	\section{Covariate Balancing}
	\label{sec:balancing}
	
	In a general missing data setting, \cite{EGP2008} and \cite{GPE2012,GPE2016} introduced inverse probability tilting (IPT) as a method for estimating the propensity score, along with other parameters of interest. In this section, we review the estimation problems solved by this method in the binary treatment case.
	
	Let $W$ denote the binary treatment indicator, and define the propensity score as $\pr ( W=1 | \mathbf{X} = \mathbf{x} )$. Assume an index model, $p ( \mathbf{x\gamma} )$, where $\mathbf{x}$ is $1 \times K$, $\mathbf{\gamma}$ is $K \times 1$, and $x_{1} \equiv 1$ ensures an intercept. Although $p ( \mathbf{x\gamma} )$ is typically taken to be logit, our results apply more generally to any index model, including probit, complementary log-log, linear, and inverse linear models. Let $Y(0)$ and $Y(1)$ denote the potential outcomes. Recall that the ATE and ATT are defined as
	\begin{equation*}
		\label{eq:ate}
		\tau_{ate} \; = \; \e \left[ Y(1) - Y(0) \right]
	\end{equation*}
	and
	\begin{equation*}
		\label{eq:att}
		\tau_{att} \; = \; \e \left[ Y(1) - Y(0) | W=1 \right].
	\end{equation*}
	However, we emphasize that the results in this paper pertain to algebraic equivalences, and therefore, we do not discuss the identification of population parameters.
	
	To fix ideas, consider the case where $p ( \mathbf{x\gamma} )$ is logit, i.e., $p ( \mathbf{x\gamma} ) = \exp ( \mathbf{x\gamma} ) / \left[ 1 + \exp ( \mathbf{x\gamma} ) \right]$. In practice, the logit model is often conflated with its estimation via maximum likelihood, in which case, for a sample of size $N$, the maximum likelihood estimator $\mathbf{\hat{\gamma}}_{mle}$ solves the first-order condition:
	\begin{equation}
		\label{eq:logfocs}
		\sum_{i=1}^{N} \mathbf{X}_{i}^{\prime} \left[ W_{i} - p ( \mathbf{X}_{i} \mathbf{\hat{\gamma}}_{mle} ) \right] \; = \; \mathbf{0}.
	\end{equation}
	IPT replaces this condition with a different set of moment equations for estimating $\mathbf{\gamma}$. While our discussion below is not limited to the logit case, the point is that even when the logit model is used, estimation need not rely on maximum likelihood; it can proceed via the method of moments instead. When $W$ is a treatment indicator, the IPT moment conditions proposed by \cite{EGP2008} and \cite{GPE2012} for estimating $\e [ Y(1) ]$ are
	\begin{equation}
		\label{eq:gpemoments}
		\e \left[ \frac{W}{p ( \mathbf{X\gamma} )} \mathbf{X}^{\prime} \right] \; = \; \e ( \mathbf{X}^{\prime} ),
	\end{equation}
	which follow immediately by iterated expectations when $p ( \mathbf{X\gamma} ) = \pr ( W=1 | \mathbf{X} ) = \e ( W | \mathbf{X} )$. (If we were considering identification, we would need to assume, at a minimum, that $p ( \mathbf{X\gamma} ) >0$ with probability one.) The sample analog of \eqref{eq:gpemoments} is
	\begin{equation}
		\label{eq:gpemoments_sample}
		N^{-1} \sum_{i=1}^{N} \frac{W_{i} \mathbf{X}_{i}}{p ( \mathbf{X}_{i} \mathbf{\hat{\gamma}}_{1,ipt} )} \; = \; \mathbf{\bar{X}},
	\end{equation}
	and these equations define the IPT estimator of $\mathbf{\gamma}$, $\mathbf{\hat{\gamma}}_{1,ipt}$, regardless of the specific model chosen for $\pr ( W=1 | \mathbf{X} = \mathbf{x} )$. Note that we have put a ``1'' subscript on $\mathbf{\hat{\gamma}}_{1,ipt}$ because, in the treatment effects setting, there is another set of moment conditions for estimating $\e [ Y(0) ]$ that leads to a different IPT estimator of $\mathbf{\gamma}$. Again, by iterated expectations,
	\begin{equation}
		\label{eq:gpemoments0}
		\e \left[ \frac{1-W}{1 - p ( \mathbf{X\gamma} )} \mathbf{X}^{\prime} \right] \; = \; \e ( \mathbf{X}^{\prime} ),
	\end{equation}
	and this leads to the sample analog:
	\begin{equation}
		\label{eq:gpemoments0_sample}
		N^{-1} \sum_{i=1}^{N} \frac{\left( 1-W_{i} \right) \mathbf{X}_{i}}{1 - p ( \mathbf{X}_{i} \mathbf{\hat{\gamma}}_{0,ipt} )} \; = \; \mathbf{\bar{X}}.
	\end{equation}
	In general, $\mathbf{\hat{\gamma}}_{0,ipt} \neq \mathbf{\hat{\gamma}}_{1,ipt}$. However, because $1 \in \mathbf{X}_{i}$, it follows immediately that
	\begin{equation}
		\label{eq:iptnorm1}
		N^{-1} \sum_{i=1}^{N} \frac{W_{i}}{p ( \mathbf{X}_{i} \mathbf{\hat{\gamma}}_{1,ipt} )} \; = \; 1
	\end{equation}
	and
	\begin{equation}
		\label{eq:iptnorm0}
		N^{-1} \sum_{i=1}^{N} \frac{1-W_{i}}{1 - p ( \mathbf{X}_{i} \mathbf{\hat{\gamma}}_{0,ipt} )} \; = \; 1.
	\end{equation}
	These two equations are key, as the summands in \eqref{eq:iptnorm1} are the weights for estimating $\e [ Y(1) ]$ in IPW estimation, and those in \eqref{eq:iptnorm0} are the weights used in estimating $\e [ Y(0) ]$, as we review in Section \ref{sec:equivalence}. Equations \eqref{eq:iptnorm1} and \eqref{eq:iptnorm0} show that the IPT weights are automatically normalized for estimating the ATE\@. That is, the sample mean of the weights is not stochastic but instead equal to one by construction. These equations also indicate that the IPT estimator of the ATE will require estimating the propensity score twice, with one set of predicted probabilities used to estimate $\e [ Y(1) ]$ and another to estimate $\e [ Y(0) ]$.
	
	For estimating the ATT, the moment equations used by \cite{EGP2008} and \cite{GPE2016} are
	\begin{equation*}
		\label{eq:irmoments_att}
		\e ( \mathbf{X}^{\prime} | W=1 ) \; = \; \frac{1}{\rho} \cdot \e ( W \cdot \mathbf{X}^{\prime} ) \; = \; \frac{1}{\rho} \cdot \e \left[ \frac{p ( \mathbf{X\gamma} ) \left( 1-W \right)}{1 - p ( \mathbf{X\gamma} )} \cdot \mathbf{X}^{\prime} \right],
	\end{equation*}
	where $\rho = \pr ( W=1 )$. Using $\hat{\rho} = N_{1}/N$, where $N_{1}$ is the number of treated units, the $K$ sample moment conditions are
	\begin{eqnarray*}
		N_{1}^{-1} \sum_{i=1}^{N} W_{i} \mathbf{X}_{i}^{\prime} &=& \left( \frac{N_{1}}{N} \right)^{-1} N^{-1} \sum_{i=1}^{N} \frac{p ( \mathbf{X}_{i} \mathbf{\hat{\gamma}}_{0,ipt} ) \left( 1-W_{i} \right)}{1 - p ( \mathbf{X}_{i} \mathbf{\hat{\gamma}}_{0,ipt} )} \cdot \mathbf{X}_{i}^{\prime} \\
		&=& N_{1}^{-1} \sum_{i=1}^{N} \frac{p ( \mathbf{X}_{i} \mathbf{\hat{\gamma}}_{0,ipt} ) \left( 1-W_{i} \right)}{1 - p ( \mathbf{X}_{i} \mathbf{\hat{\gamma}}_{0,ipt} )} \cdot \mathbf{X}_{i}^{\prime}
	\end{eqnarray*}
	or
	\begin{equation}
		\label{eq:irmoments_att_sample}
		\mathbf{\bar{X}}_{1}^{\prime} \; = \; N_{1}^{-1} \sum_{i=1}^{N} \frac{p ( \mathbf{X}_{i} \mathbf{\hat{\gamma}}_{0,ipt} ) \left( 1-W_{i} \right)}{1 - p ( \mathbf{X}_{i} \mathbf{\hat{\gamma}}_{0,ipt} )} \cdot \mathbf{X}_{i}^{\prime},
	\end{equation}
	where $\mathbf{\bar{X}}_{1} = N_{1}^{-1} \sum_{i=1}^{N} W_{i} \mathbf{X}_{i}$\@. Because $1 \in \mathbf{X}_{i}$, \eqref{eq:irmoments_att_sample} implies
	\begin{equation*}
		N_{1} \; = \; \sum_{i=1}^{N} \frac{p ( \mathbf{X}_{i} \mathbf{\hat{\gamma}}_{0,ipt} ) \left( 1-W_{i} \right)}{1 - p ( \mathbf{X}_{i} \mathbf{\hat{\gamma}}_{0,ipt} )},
	\end{equation*}	
	which implies that the weights used in the IPW estimation of the ATT sum to the number of treated units. In other words, like in the case of the ATE, the IPT weights for estimating the ATT are automatically normalized.
	
	It may seem surprising that we use $\mathbf{\hat{\gamma}}_{0,ipt}$ to denote the IPT estimator of $\mathbf{\gamma}$ in the context of estimating the ATT, given that we used the same notation in equations \eqref{eq:gpemoments0_sample} and \eqref{eq:iptnorm0} above. This is fully warranted, however, because this estimator is in fact the same as the IPT estimator defined by equation \eqref{eq:gpemoments0_sample}. Indeed, as shown by \cite{Tan2020}, the moment conditions in \eqref{eq:gpemoments0_sample}, which balance the weighted covariates of the comparison group with those of the overall sample, are algebraically equivalent to the conditions in \eqref{eq:irmoments_att_sample}, which instead balance them with the treated group, but using a different set of weights. To see this equivalence, we can rewrite the sample moment conditions in \eqref{eq:gpemoments0_sample} as
	\begin{equation*}
		N^{-1} \sum_{i=1}^{N} \left( \frac{ 1-W_{i} }{1 - p ( \mathbf{X}_{i} \mathbf{\hat{\gamma}}_{0,ipt} )} - 1 \right) \cdot \mathbf{X}_{i}^{\prime} \; = \; \mathbf{0},
	\end{equation*}
	which, after simple algebra, can be expressed as
	\begin{equation*}
		N^{-1} \sum_{i=1}^{N} \frac{p ( \mathbf{X}_{i} \mathbf{\hat{\gamma}}_{0,ipt} ) \left( 1-W_{i} \right)}{1 - p ( \mathbf{X}_{i} \mathbf{\hat{\gamma}}_{0,ipt} )} \cdot \mathbf{X}_{i}^{\prime} \; = \; N^{-1} \sum_{i=1}^{N} W_{i} \mathbf{X}_{i}^{\prime},
	\end{equation*}
	and this, in turn, is easily seen as equivalent to equation \eqref{eq:irmoments_att_sample}.
	
	It is also useful to briefly compare the IPT moment conditions for estimating the ATE with a subsequent proposal by \cite{IR2014}, known as the ``covariate balancing propensity score (CBPS),'' which uses different moment conditions to obtain a single estimator of $\mathbf{\gamma}$. Again, if the propensity score is correctly specified then, by iterated expectations,
	\begin{equation}
		\label{eq:irmoments}
		\e ( \mathbf{X}^{\prime} ) \; = \; \e \left[ \frac{W}{p ( \mathbf{X\gamma} )} \mathbf{X}^{\prime} \right] \; = \; \e \left[ \frac{1-W}{1 - p ( \mathbf{X\gamma} )} \mathbf{X}^{\prime} \right].
	\end{equation}
	Rather than using the implications of \eqref{eq:irmoments} separately, which is what IPT does, \cite{IR2014} use the second equality to obtain the following sample moment conditions:
	\begin{equation}
		\label{eq:irmoments_sample_1}
		N^{-1} \sum_{i=1}^{N} \frac{W_{i}}{p ( \mathbf{X}_{i} \mathbf{\hat{\gamma}}_{cbps} )} \mathbf{X}_{i}^{\prime} \; = \; N^{-1} \sum_{i=1}^{N} \frac{1-W_{i}}{1 - p ( \mathbf{X}_{i} \mathbf{\hat{\gamma}}_{cbps} )} \mathbf{X}_{i}^{\prime}.
	\end{equation}
	After simple algebra, the moment conditions can be expressed as
	\begin{equation}
		\label{eq:irmoments_sample_2}
		\sum_{i=1}^{N} \left( \frac{W_{i} - p ( \mathbf{X}_{i} \mathbf{\hat{\gamma}}_{cbps} )}{p ( \mathbf{X}_{i} \mathbf{\hat{\gamma}}_{cbps} ) \left[ 1 - p ( \mathbf{X}_{i} \mathbf{\hat{\gamma}}_{cbps} ) \right]} \right) \mathbf{X}_{i}^{\prime} \; = \; \mathbf{0}.
	\end{equation}
	Comparing \eqref{eq:irmoments_sample_2} with \eqref{eq:logfocs}, we can see that the CBPS approach---when applied to the logit model---weights the MLE moment conditions by the estimated inverse conditional variance, $\mathrm{Var} ( W_{i} | \mathbf{X}_{i} )$.
	
	Because the first element of $\mathbf{X}_{i}$ is unity, \eqref{eq:irmoments_sample_1} also implies that
	\begin{equation}
		\label{eq:cbpsunnorm}
		\sum_{i=1}^{N} \frac{W_{i}}{p ( \mathbf{X}_{i} \mathbf{\hat{\gamma}}_{cbps} )} \; = \; \sum_{i=1}^{N} \frac{1-W_{i}}{1 - p ( \mathbf{X}_{i} \mathbf{\hat{\gamma}}_{cbps} )}.
	\end{equation}
	Equation \eqref{eq:cbpsunnorm} shows that the weights appearing in the IPW estimates of $\e [ Y(1) ]$ and $\e [ Y(0) ]$ sum to the same value, but that common value is not necessarily the sample size, $N$. In other words, when these are used as weights in IPW, the CBPS weights are not automatically normalized.
	
	Finally, when estimating the ATT, \cite{IR2014} suggest using the moment conditions in equation \eqref{eq:irmoments_att_sample}, following the approach of \cite{EGP2008} and \cite{GPE2016}. This implies that the IPT and CBPS estimators of the ATT are the same. When using the logit model, as shown by \cite{Tan2020}, both approaches are also numerically identical to the entropy balancing estimator of \cite{Hainmueller2012}.\footnote{The three estimators of the ATT will no longer coincide if, in the case of CBPS, the IPT moment conditions are combined with the first-order conditions of the maximum likelihood estimator (the so-called ``overidentified CBPS'')\@. Likewise, the entropy balancing estimator will differ from IPT and CBPS if its implementation constrains higher moments of the covariates to be balanced, too.}

	\section{Equivalence of Estimators}
	\label{sec:equivalence}
	
	In this section, we establish numerical equivalences among three different classes of estimators that incorporate inverse probability weighting, starting with estimators of the ATE.
	
	\subsection{Estimators of the ATE}
	\label{sec:ATE}
	
	The three estimators we consider are among the most popular alternatives to OLS estimation of an additive linear model when unconfoundedness is assumed to hold: IPW, AIPW, and IPWRA\@. As we establish algebraic equivalence, we do not impose assumptions beyond those necessary for the existence of estimates for a given sample. This simply means that the estimated propensity scores are strictly between zero and one for all $i$.
	
	The IPW estimator of $\tau _{ate}$ using the IPT weights is
	\begin{equation}
		\label{eq:ipt_ate}
		\hat{\tau}_{ate,ipt} \; = \; \hat{\mu}_{1,ipt} \; - \; \hat{\mu}_{0,ipt} \; = \; N^{-1} \sum_{i=1}^{N} \frac{W_{i}Y_{i}}{p ( \mathbf{X}_{i} \mathbf{\hat{\gamma}}_{1,ipt} ) } \; - \; N^{-1} \sum_{i=1}^{N} \frac{\left( 1-W_{i} \right) Y_{i}}{1 - p ( \mathbf{X}_{i} \mathbf{\hat{\gamma}}_{0,ipt} )},
	\end{equation}
	where the subscript ``ipt'' indicates the use of IPT weights. See, e.g., \citet[Section 21.3]{Wooldridge2010} for a variant of this estimator with MLE-based weights and \cite{EGP2008} and \cite{GPE2012} for IPT\@. We know from \eqref{eq:iptnorm1} and \eqref{eq:iptnorm0} that the weights in both weighted averages are automatically normalized.
	
	The AIPW estimator with IPT weights, which we refer to as AIPT, is also the difference in estimates of $\mu_{1} \equiv \e [ Y(1) ]$ and $\mu_{0} \equiv \e [ Y(0) ]$; that is, $\hat{\tau}_{ate,aipt} = \hat{\mu}_{1,aipt} - \hat{\mu}_{0,aipt}$. For $\mu_{1}$,
	\begin{equation}
		\label{eq:aipt_mu1}
		\hat{\mu}_{1,aipt} \; = \; N^{-1} \sum_{i=1}^{N} \frac{W_{i} \left( Y_{i} - \mathbf{X}_{i} \mathbf{\hat{\beta}}_{1} \right)}{p ( \mathbf{X}_{i} \mathbf{\hat{\gamma}}_{1,ipt} )} \; + \; N^{-1} \sum_{i=1}^{N} \mathbf{X}_{i} \mathbf{\hat{\beta}}_{1},
	\end{equation}
	where remember that $1 \in \mathbf{X}_{i}$. Although it is not important for the equivalence result, the estimates $\mathbf{\hat{\beta}}_{1}$ typically come from an OLS regression of $Y_{i}$ on $\mathbf{X}_{i}$ using $W_{i}=1$ (treated units). The first term in \eqref{eq:aipt_mu1} is a weighted average of the resulting residuals over the treated units. The weights are exactly those appearing in $\hat{\mu}_{1,ipt}$ and are therefore normalized.\footnote{Normalization is less important in AIPW than in IPW\@. \cite{Knaus2024} shows that ``unnormalized'' AIPW, unlike unnormalized IPW, can still be expressed as a weighted average of observed outcomes with weights that sum to one. This normalization is automatic under standard implementations of outcome regressions.}
	
	For $\mu_{0}$, the AIPT estimator is
	\begin{equation}
		\label{eq:aipt_mu0}
		\hat{\mu}_{0,aipt} \; = \; N^{-1} \sum_{i=1}^{N} \frac{\left( 1-W_{i} \right) \left( Y_{i} - \mathbf{X}_{i} \mathbf{\hat{\beta}}_{0} \right)}{1 - p ( \mathbf{X}_{i} \mathbf{\hat{\gamma}}_{0,ipt} )} \; + \; N^{-1} \sum_{i=1}^{N} \mathbf{X}_{i} \mathbf{\hat{\beta}}_{0},
	\end{equation}
	where $\mathbf{\hat{\beta}}_{0}$ are probably the OLS estimates from a regression of $Y_{i}$ on $\mathbf{X}_{i}$ using $W_{i}=0$.
	
	The third estimator we consider is the IPWRA estimator with IPT weights, which we refer to as IPTRA\@. For $\mu_{1}$, we first solve a weighted least squares (WLS) problem,
	\begin{equation}
		\label{eq:problem_wls1}
		\min_{\mathbf{b}_{1}} \text{ } N^{-1} \sum_{i=1}^{N} \frac{W_{i}}{\hat{p}_{i}} \left( Y_{i} - \mathbf{X}_{i} \mathbf{b}_{1} \right)^{2},
	\end{equation}
	where $\hat{p}_{i} = p ( \mathbf{X}_{i} \mathbf{\hat{\gamma}}_{1,ipt} )$ are the IPT propensity score estimates. Given the WLS estimates $\mathbf{\tilde{\beta}}_{1}$ from \eqref{eq:problem_wls1}, $\mu_{1}$ is estimated by averaging the fitted values across all observations, as in the case of linear regression adjustment:
	\begin{equation}
		\label{eq:iptra_mu1}
		\hat{\mu}_{1,iptra} \; = \; N^{-1} \sum_{i=1}^{N} \mathbf{X}_{i} \mathbf{\tilde{\beta}}_{1} \; = \; \mathbf{\bar{X}\tilde{\beta}}_{1}.
	\end{equation}
	The IPTRA estimator of $\mu_{0}$, $\hat{\mu}_{0,iptra}$, uses the untreated units with weights $\left( 1-p ( \mathbf{X}_{i} \mathbf{\hat{\gamma}}_{0,ipt} ) \right)^{-1}$, and produces $\mathbf{\tilde{\beta}}_{0}$. The final IPTRA estimator of the ATE is given by $\hat{\tau}_{ate,iptra} = \hat{\mu}_{1,iptra} - \hat{\mu}_{0,iptra} = \mathbf{\bar{X}\tilde{\beta}}_{1} - \mathbf{\bar{X}\tilde{\beta}}_{0}$.
	
	When the inverse probability weights are obtained using MLE, CBPS, or some other method of moments procedure, $\hat{\tau}_{ate,ipw}$, $\hat{\tau}_{ate,aipw}$, and $\hat{\tau}_{ate,ipwra}$ are generally different. In fact, $\hat{\tau}_{ate,ipw}$ and $\hat{\tau}_{ate,aipw}$ do not generally use normalized weights, and so one could have five different estimates using the same estimated weights: IPW, normalized IPW (NIPW), AIPW, normalized AIPW (NAIPW), and IPWRA\@. (IPWRA is always normalized.) Strikingly, when IPT weights are used instead, all of these estimates are identical.
	\medskip
	\begin{proposition}
		\label{prop:ATE}
		Let $\mathbf{\hat{\gamma}}_{1,ipt}$ be the estimates from the IPT estimation in equation \eqref{eq:gpemoments_sample}, with $\hat{p}_{i} = p ( \mathbf{X}_{i} \mathbf{\hat{\gamma}}_{1,ipt} ) > 0$ for all $i$. Then $\hat{\mu}_{1,ipt}$, $\hat{\mu}_{1,aipt}$, and $\hat{\mu}_{1,iptra}$ are numerically identical. The same is true of $\hat{\mu}_{0,ipt}$, $\hat{\mu}_{0,aipt}$, and $\hat{\mu}_{0,iptra}$, which means that
		\begin{equation*}
			\hat{\tau}_{ate,ipt} \; = \; \hat{\tau}_{ate,aipt} \; = \; \hat{\tau}_{ate,iptra}.
		\end{equation*}
	\end{proposition}
	\medskip
	\noindent
	The implication of Proposition \ref{prop:ATE} is that if one uses the IPT weights in estimating both $\mu_{0}$ and $\mu_{1}$, where conditional means $\e [ Y(0) | \mathbf{X} ]$ and $\e [ Y(1) | \mathbf{X} ]$ are modeled linearly, then three prominent estimators of the ATE are numerically identical; moreover, the IPW and AIPW versions are automatically normalized.

	\subsection{Estimators of the ATT}
	\label{sec:ATT}
	
	We now establish the equivalence of several prominent estimators of the ATT when the IPT weights from equation \eqref{eq:irmoments_att_sample} are used. Recall that
	\begin{equation*}
		\tau_{att} \; = \; \e [ Y(1) | W=1 ] - \e [ Y(0) | W=1 ] \; \equiv \; \mu_{1|1} - \mu_{0|1},
	\end{equation*}
	and the first term is always consistently estimated using the sample mean of $Y_{i}$ over the treated units, $\bar{Y}_{1}$. The IPW estimator for the second term, using the IPT weights, is
	\begin{equation}
		\label{eq:ipwcbps_mu01}
		\hat{\mu}_{0|1,ipt} \; = \; N_{1}^{-1} \sum_{i=1}^{N} \frac{p ( \mathbf{X}_{i} \mathbf{\hat{\gamma}}_{0,ipt} ) \left( 1-W_{i} \right) Y_{i}}{1 - p ( \mathbf{X}_{i} \mathbf{\hat{\gamma}}_{0,ipt} )}.
	\end{equation}
	As noted earlier, the weights in \eqref{eq:ipwcbps_mu01} sum to the number of treated units; thus, they are automatically normalized. The same weights also appear in the AIPW estimator. Therefore, the normalized and unnormalized IPW estimators coincide, as do the normalized and unnormalized AIPW estimators. Specifically, the AIPW estimator of $\mu _{0|1}$ is
	\begingroup
	\allowdisplaybreaks
	\begin{eqnarray}
		\hat{\mu}_{0|1,aipt} &=& N_{1}^{-1} \sum_{i=1}^{N} \frac{\hat{p}_{i} \left( 1-W_{i} \right)}{1 - \hat{p}_{i}} \left( Y_{i} - \mathbf{X}_{i} \mathbf{\hat{\beta}}_{0} \right) \; + \; N_{1}^{-1} \sum_{i=1}^{N} W_{i} \mathbf{X}_{i} \mathbf{\hat{\beta}}_{0} \nonumber \\
		\label{eq:aipwcbps_mu01}
		&=& N_{1}^{-1} \sum_{i=1}^{N} \frac{\hat{p}_{i} \left( 1-W_{i} \right) Y_{i}}{1 - \hat{p}_{i}} \; - \; N_{1}^{-1} \sum_{i=1}^{N} \frac{\hat{p}_{i} \left( 1-W_{i} \right) \mathbf{X}_{i} \mathbf{\hat{\beta}}_{0}}{1 - \hat{p}_{i}} \; + \; \mathbf{\bar{X}}_{1} \mathbf{\hat{\beta}}_{0},
	\end{eqnarray}
	\endgroup
	where $\hat{p}_{i} = p ( \mathbf{X}_{i} \mathbf{\hat{\gamma}}_{0,ipt} )$ are now the IPT propensity score estimates and $\mathbf{\hat{\beta}}_{0}$ is typically the OLS estimator from regressing $Y_{i}$ on $\mathbf{X}_{i}$ using $W_{i}=0$.
	
	Finally, the IPWRA estimator of $\mu _{0|1}$, using the weights from \eqref{eq:irmoments_att_sample}, is
	\begin{equation*}
		\hat{\mu}_{0|1,iptra} \; = \; \mathbf{\bar{X}}_{1} \mathbf{\tilde{\beta}}_{0},
	\end{equation*}
	where $\mathbf{\tilde{\beta}}_{0}$ now solves the WLS problem:
	\begin{equation}
		\label{eq:problem_wls0}
		\min_{\mathbf{b}_{0}} \text{ } N^{-1} \sum_{i=1}^{N} \frac{\hat{p}_{i} \left( 1-W_{i} \right)}{1 - \hat{p}_{i}} \left( Y_{i} - \mathbf{X}_{i} \mathbf{b}_{0} \right)^{2},
	\end{equation}
	where $\hat{p}_{i} = p ( \mathbf{X}_{i} \mathbf{\hat{\gamma}}_{0,ipt} )$. We have the following equivalence result.
	\medskip
	\begin{proposition}
		\label{prop:ATT}
		Let $\mathbf{\hat{\gamma}}_{0,ipt}$ be the estimators solving \eqref{eq:irmoments_att_sample} with $\hat{p}_{i} = p ( \mathbf{X}_{i} \mathbf{\hat{\gamma}}_{0,ipt} ) < 1$ for all $i$. Then, the IPW, AIPW, and IPWRA estimates of $\mu_{0|1}$, using the IPT weights and linear conditional means in the latter two cases, are identical. Therefore, the three estimates of $\tau_{att}$ are identical.
	\end{proposition}
	\medskip
	\noindent
	Similar to Proposition \ref{prop:ATE}, the implication of Proposition \ref{prop:ATT} is that if one uses the IPT weights in estimating $\mu_{0|1}$ as well as a linear model for $\e [ Y(0) | \mathbf{X} ]$, then the IPW, AIPW, and IPWRA estimators of the ATT are numerically identical; moreover, the IPW and AIPW versions are automatically normalized.

	\section{Implications for Instrumental Variables and Difference-in-Differences Settings}
	\label{sec:implications}
	
	In this section, we briefly discuss the implications of the results in Section \ref{sec:equivalence} for estimating local average treatment effects with instrumental variables and heterogeneous treatment effects in difference-in-differences settings.
	
	\subsection{Instrumental Variables}
	\label{sec:LATE}
	
	The results in Section \ref{sec:equivalence} have implications for estimators of the local average treatment effect (LATE) and the local average treatment effect on the treated (LATT) when using control variables $\mathbf{X}$; a recent treatment is \cite{SUW2022}, which we follow here. As before, $W$ is a treatment variable. We assume it to be binary, although this can be easily relaxed. We also have a binary instrumental variable, $Z$.
	
	It follows from \cite{Frolich2007} that many estimators of the LATE are ratios of estimators of the ATE,
	\begin{equation*}
		\hat{\tau}_{late} \; = \; \frac{\hat{\tau}_{ate,Y|Z}}{\hat{\tau}_{ate,W|Z}},
	\end{equation*}
	where $\hat{\tau}_{ate,Y|Z}$ is an estimator of the ATE where $Y$ is the outcome, $Z$ plays the role of the treatment, and the covariates $\mathbf{X}$ are used to account for confounders of $Z$\@. Again, we are only concerned with equivalences and not statistical properties. The denominator, $\hat{\tau}_{ate,W|Z}$, is an estimated ATE where $W$ is the outcome and $Z$ again is the treatment indicator, with covariates $\mathbf{X}$. It follows from Proposition \ref{prop:ATE} that when linear conditional means are used for both $Y$ and $W$, and IPT is used for the weights, estimators of the LATE based on IPW, AIPW, and IPWRA are all identical. The inverse probability weights, in this case, for both the numerator and the denominator, are based on the instrument propensity score:
	\begin{equation*}
		\pr ( Z=1 | \mathbf{X} ) \; = \; q ( \mathbf{X\delta} ).
	\end{equation*}
	It should be noted, however, that unlike in the case of the ATE, where the nature of $Y$ is generally unspecified, here it may be impractical to use the linear model in the denominator when $W$ is binary. See \cite{SUW2022} for using other doubly robust estimators to exploit the binary nature of $W$ and maybe special features of $Y$. In such cases, however, the numerical equivalence results no longer hold.
	
	Estimators of the LATT that incorporate control variables $\mathbf{X}$ can be written as the ratio of estimators of the ATT, where the instrument plays the role of the treatment variable:
	\begin{equation*}
		\hat{\tau}_{latt} \; = \; \frac{\hat{\tau}_{att,Y|Z}}{\hat{\tau}_{att,W|Z}},
	\end{equation*}
	where $\hat{\tau}_{att,Y|Z}$ and $\hat{\tau}_{att,W|Z}$ are both estimators of the ATT with ``treatment'' variable $Z$ and outcome variables $Y$ and $W$, respectively. If these estimators use the appropriate IPT weights, as in Proposition \ref{prop:ATT}, then it follows immediately that the estimators of $\tau_{latt}$ based on IPW, AIPW with linear regression functions, and IPWRA with linear regression functions are all numerically the same. Also, recall that the normalized and unnormalized estimators of the ATT are identical when using these weights.

	\subsection{Difference-in-Differences}
	\label{sec:DID}
	
	Some popular estimators in difference-in-differences (DID) settings are based on applying standard treatment effect estimators after suitably transforming the outcome variable. For example, \cite{Abadie2005}, with two time periods, proposes applying IPW to the differences $Y_{i2}-Y_{i1}$, where $Y_{it}$ is the outcome for unit $i$ in period $t$. \cite{Abadie2005} uses maximum likelihood estimation of the propensity score to construct the inverse probability weights. \cite{SAZ2020} instead develop a doubly robust estimator of the ATT\@. The estimator uses a structure similar to equation \eqref{eq:aipwcbps_mu01}, although \cite{SAZ2020} also normalize the weights and replace the OLS estimates of the conditional mean of the untreated potential change in outcomes with the WLS estimates similar to equation \eqref{eq:problem_wls0}; they also use the IPT weights from equation \eqref{eq:irmoments_att_sample}. Strikingly, the results in Section \ref{sec:equivalence} imply that all these additional modifications have no impact on the final estimate of the ATT when the IPT weights are used; in other words, the simple IPW estimator in \cite{Abadie2005} is numerically identical to \cite{SAZ2020} when both use the IPT moment conditions to estimate the propensity score.
	
	Similar conclusions hold with many periods and staggered interventions. \cite{CSA2021} extend \cite{SAZ2020} to estimate ATTs by treatment cohort (i.e., the first period of treatment), $g$, and calendar time, $t$. To estimate these ATTs, $\tau_{gt}$, \cite{CSA2021} apply different versions of AIPW and IPWRA to differences $Y_{it}-Y_{i,g-1}$, where $Y_{i,g-1}$ is the outcome in the period just before the first treatment period for treatment cohort $g$. \cite{CSA2021} emphasize that the comparison group can either consist of the never treated (NT) cohort or the NT cohort plus other cohorts that are first treated in period $t+1$ or later (``not yet treated''). One of the estimators recommended by \cite{CSA2021} is AIPW with the IPT weights from equation \eqref{eq:irmoments_att_sample}. It follows immediately that applying IPW, AIPW, or IPWRA to $Y_{it}-Y_{i,g-1}$ with treatment indicator $D_{ig}$ (indicating treatment cohort) and controls $\mathbf{X}_{i}$ delivers identical estimates of the $\tau_{gt}$ when IPT weights are used. This is true even when $t<g-1$, which provides event study graphs for studying the existence of pre-trends.
	
	An alternative transformation in the staggered intervention settings uses data on all pre-treatment outcomes by removing the average of the outcomes over all pre-treatment periods: $\dot{Y}_{itg} \equiv Y_{it} - \left( g-1 \right) ^{-1} \sum_{s=1}^{g-1} Y_{is}$. As shown in \cite{LW2024}, under standard no anticipation and conditional parallel trends assumptions, one can apply various treatment effect estimators to the cross-sectional data $\left\{ \left( \dot{Y}_{itg},D_{ig},\mathbf{X}_{i}\right) :i=1,...,N\right\} $ to consistently estimate $\tau _{gt}$. Again, the results in Section \ref{sec:equivalence} immediately imply that the IPW, AIPW, and IPWRA estimators with the IPT weights from equation \eqref{eq:irmoments_att_sample} are all identical when applied to these data once one chooses a suitable comparison group.

	\section{Empirical Applications}
	\label{sec:applications}
	
	In this section, we illustrate our findings with two empirical applications, beginning with a replication of a prominent study of the causal effects of cash transfers on longevity \citep{AEFLM2016} and concluding with a reanalysis of the empirical application in \cite{SAZ2020}.
	
	\subsection{The Effects of Cash Transfers on Longevity}
	\label{sec:aizer}
	
	\begin{table}[!p]
		\begin{adjustwidth}{-1.25in}{-1.25in}
			\centering
			\begin{threeparttable}
				\caption{Replication of \cite{AEFLM2016}\label{tab:aizer}}
				\begin{footnotesize}
					\begin{tabular}{c >{\centering\arraybackslash}m{1.4cm} >{\centering\arraybackslash}m{1.4cm} >{\centering\arraybackslash}m{0.01cm} >{\centering\arraybackslash}m{1.4cm} >{\centering\arraybackslash}m{1.4cm} >{\centering\arraybackslash}m{0.01cm} >{\centering\arraybackslash}m{1.4cm} >{\centering\arraybackslash}m{1.4cm} >{\centering\arraybackslash}m{0.01cm} >{\centering\arraybackslash}m{1.4cm} >{\centering\arraybackslash}m{1.4cm}}
						\hline\hline
						& \multicolumn{2}{c}{(1)} &       & \multicolumn{2}{c}{(2)} &       & \multicolumn{2}{c}{(3)} &       & \multicolumn{2}{c}{(4)} \\
						\hline
						OLS   & \multicolumn{2}{c}{0.0157***} &       & \multicolumn{2}{c}{0.0158***} &       & \multicolumn{2}{c}{0.0163***} &       & \multicolumn{2}{c}{0.0146**} \\
						& \multicolumn{2}{c}{(0.0058)} &       & \multicolumn{2}{c}{(0.0059)} &       & \multicolumn{2}{c}{(0.0059)} &       & \multicolumn{2}{c}{(0.0059)} \\
						\hline
						& ATE   & ATT   &       & ATE   & ATT   &       & ATE   & ATT   &       & ATE   & ATT \\
						\cline{2-3} \cline{5-6} \cline{8-9} \cline{11-12}
						RA    & 0.0105* & 0.0096 &       & 0.0100 & 0.0092 &       & 0.0100 & 0.0090 &       & 0.0080 & 0.0069 \\
						& (0.0062) & (0.0063) &       & (0.0068) & (0.0071) &       & (0.0071) & (0.0075) &       & (0.0071) & (0.0075) \\
						\hline
						& \multicolumn{11}{c}{MLE weights} \\
						\cline{2-12}
						& ATE   & ATT   &       & ATE   & ATT   &       & ATE   & ATT   &       & ATE   & ATT \\
						\cline{2-3} \cline{5-6} \cline{8-9} \cline{11-12}
						IPW   & 0.0597*** & 0.0645*** &       & 0.0438 & 0.0391 &       & 0.0113 & 0.0002 &       & 0.0093 & --0.0014 \\
						& (0.0164) & (0.0181) &       & (0.0886) & (0.1002) &       & (0.1216) & (0.1381) &       & (0.1222) & (0.1387) \\
						NIPW  & 0.0107* & 0.0099 &       & 0.0064 & 0.0047 &       & 0.0025 & 0.0001 &       & 0.0014 & --0.0006 \\
						& (0.0061) & (0.0063) &       & (0.0072) & (0.0077) &       & (0.0072) & (0.0077) &       & (0.0072) & (0.0076) \\
						AIPW  & 0.0102* & 0.0093 &       & 0.0056 & 0.0040 &       & 0.0042 & 0.0023 &       & 0.0024 & 0.0007 \\
						& (0.0062) & (0.0064) &       & (0.0069) & (0.0073) &       & (0.0072) & (0.0076) &       & (0.0072) & (0.0077) \\
						NAIPW & 0.0102* & 0.0093 &       & 0.0056 & 0.0039 &       & 0.0042 & 0.0023 &       & 0.0024 & 0.0007 \\
						& (0.0062) & (0.0064) &       & (0.0069) & (0.0073) &       & (0.0072) & (0.0076) &       & (0.0072) & (0.0076) \\
						IPWRA & 0.0101 & 0.0092 &       & 0.0072 & 0.0058 &       & 0.0050 & 0.0028 &       & 0.0037 & 0.0019 \\
						& (0.0062) & (0.0064) &       & (0.0066) & (0.0069) &       & (0.0069) & (0.0073) &       & (0.0069) & (0.0073) \\
						\hline
						& \multicolumn{11}{c}{IPT weights} \\
						\cline{2-12}
						& ATE   & ATT   &       & ATE   & ATT   &       & ATE   & ATT   &       & ATE   & ATT \\
						\cline{2-3} \cline{5-6} \cline{8-9} \cline{11-12}
						IPW   & 0.0101 & 0.0092 &       & 0.0076 & 0.0063 &       & 0.0057 & 0.0040 &       & 0.0047 & 0.0032 \\
						& (0.0062) & (0.0064) &       & (0.0066) & (0.0069) &       & (0.0067) & (0.0071) &       & (0.0067) & (0.0071) \\
						NIPW  & 0.0101 & 0.0092 &       & 0.0076 & 0.0063 &       & 0.0057 & 0.0040 &       & 0.0047 & 0.0032 \\
						& (0.0062) & (0.0064) &       & (0.0066) & (0.0069) &       & (0.0067) & (0.0071) &       & (0.0067) & (0.0071) \\
						AIPW  & 0.0101 & 0.0092 &       & 0.0076 & 0.0063 &       & 0.0057 & 0.0040 &       & 0.0047 & 0.0032 \\
						& (0.0062) & (0.0064) &       & (0.0066) & (0.0069) &       & (0.0067) & (0.0071) &       & (0.0067) & (0.0071) \\
						NAIPW & 0.0101 & 0.0092 &       & 0.0076 & 0.0063 &       & 0.0057 & 0.0040 &       & 0.0047 & 0.0032 \\
						& (0.0062) & (0.0064) &       & (0.0066) & (0.0069) &       & (0.0067) & (0.0071) &       & (0.0067) & (0.0071) \\
						IPWRA & 0.0101 & 0.0092 &       & 0.0076 & 0.0063 &       & 0.0057 & 0.0040 &       & 0.0047 & 0.0032 \\
						& (0.0062) & (0.0064) &       & (0.0066) & (0.0069) &       & (0.0067) & (0.0071) &       & (0.0067) & (0.0071) \\
						\hline
						State fixed effects & \multicolumn{2}{c}{\checkmark} &       & \multicolumn{2}{c}{} &       & \multicolumn{2}{c}{\checkmark} &       & \multicolumn{2}{c}{\checkmark} \\
						Cohort fixed effects & \multicolumn{2}{c}{\checkmark} &       & \multicolumn{2}{c}{\checkmark} &       & \multicolumn{2}{c}{\checkmark} &       & \multicolumn{2}{c}{\checkmark} \\
						Individual controls & \multicolumn{2}{c}{} &       & \multicolumn{2}{c}{\checkmark} &       & \multicolumn{2}{c}{\checkmark} &       & \multicolumn{2}{c}{\checkmark} \\
						State characteristics & \multicolumn{2}{c}{} &       & \multicolumn{2}{c}{\checkmark} &       & \multicolumn{2}{c}{\checkmark} &       & \multicolumn{2}{c}{\checkmark} \\
						County characteristics & \multicolumn{2}{c}{} &       & \multicolumn{2}{c}{\checkmark} &       & \multicolumn{2}{c}{\checkmark} &       & \multicolumn{2}{c}{\checkmark} \\
						&       &       &       &       &       &       &       &       &       &       &  \\
						Observations & \multicolumn{2}{c}{7,860} &       & \multicolumn{2}{c}{7,859} &       & \multicolumn{2}{c}{7,859} &       & \multicolumn{2}{c}{7,857} \\
						\hline
					\end{tabular}
				\end{footnotesize}
				\begin{footnotesize}
					\begin{tablenotes}[flushleft]
						\item \textit{Notes:} The source of data is \cite{AEFLM2016}. The outcome of interest is the log age at death, as reported in program records (specifications 1--3) or on the death certificate (specification 4). ``OLS'' corresponds to the OLS estimation of an additive linear model. ``RA'' corresponds to the regression adjustment estimator, which is based on the OLS estimation of a fully interacted linear model. The remaining estimators use a logit model for the propensity score and are defined in Section \ref{sec:equivalence}. Standard errors are in parentheses.
						\item *Statistically significant at the 10\% level; **at the 5\% level; ***at the 1\% level.
					\end{tablenotes}
				\end{footnotesize}
			\end{threeparttable}
		\end{adjustwidth}
	\end{table}
	
	\cite{AEFLM2016} study the long-run impacts of the Mothers' Pension (MP) program, which was the first government-sponsored welfare program in the prewar U.S\@. The outcome studied by the authors is the log age at death of children of the program participants. A key strength of the original study is in its careful construction of the comparison group, which consists only of mothers who were initially deemed eligible for participation but were later rejected. Still, \cite{Sloczynski2022} argues that some of the conclusions of this paper are not robust to treatment effect heterogeneity.
	
	In our application, we use the same data as \cite{AEFLM2016} and \cite{Sloczynski2022}. We consider three covariate specifications and two sources of information on dates of death: program records and death certificates. In our first specification, we control for cohort and state fixed effects. In our second specification, in line with \cite{AEFLM2016}, we replace state fixed effects with a battery of individual, county, and state characteristics. In our final specification, we reintroduce state fixed effects without dropping any other covariates.\footnote{The final specification in \cite{AEFLM2016} uses county rather than state fixed effects but is otherwise identical. In our application, using county fixed effects is not feasible because, in several counties, every eligible applicant was treated, resulting in a failure of overlap.}
	
	Table \ref{tab:aizer} reports a number of estimates of the effects of cash transfers on longevity. As in \cite{AEFLM2016}, the OLS estimates from an additive model, which controls for program participation and covariates but not interactions between the two, strongly suggest that cash transfers positively influenced the longevity of the children of their beneficiaries. However, in line with the replication in \cite{Sloczynski2022}, the majority of the estimates of the ATE and ATT are smaller or much smaller than the OLS estimates and not statistically significant. At the same time, there is a clear difference between the two panels of Table \ref{tab:aizer} that report weighting and doubly robust estimators based on MLE and IPT weights. In the case of MLE, there is a wide variation in estimates, which range from 0.0014 to 0.0597 for the ATE and from --0.0014 to 0.0645 for the ATT\@. Conditional on choosing a specific covariate specification, the choice of an estimator can have a profound impact on the researcher's conclusion. On the other hand, in the case of IPT, this choice is entirely inconsequential, as there are no differences across estimators conditional on a particular specification choice. This illustrates Propositions \ref{prop:ATE} and \ref{prop:ATT}, which demonstrate the underlying numerical equivalences. Moreover, in the case of IPT, the standard errors are usually slightly smaller than in the case of the corresponding estimates based on MLE\@.

	\subsection{The Effects of a Training Program on Earnings}
	\label{sec:lalonde}
	
	\begin{table}[!p]
		\begin{adjustwidth}{-1.25in}{-1.25in}
			\centering
			\begin{threeparttable}
				\caption{Replication of \cite{SAZ2020}\label{tab:lalonde}}
				\begin{footnotesize}
					\begin{tabular}{c >{\centering\arraybackslash}m{1.5cm} >{\centering\arraybackslash}m{1.5cm} >{\centering\arraybackslash}m{1.5cm} >{\centering\arraybackslash}m{0.01cm} >{\centering\arraybackslash}m{1.5cm} >{\centering\arraybackslash}m{1.5cm} >{\centering\arraybackslash}m{1.5cm} >{\centering\arraybackslash}m{0.01cm} >{\centering\arraybackslash}m{1.5cm} >{\centering\arraybackslash}m{1.5cm} >{\centering\arraybackslash}m{1.5cm}}
						\hline\hline
						& \multicolumn{3}{c}{LaLonde sample} &       & \multicolumn{3}{c}{DW sample} &       & \multicolumn{3}{c}{Early RA sample} \\
						\hline
						TWFE  & 868** & 868** & 868** &       & 2,092*** & 2,092*** & 2,092*** &       & 1,136  & 1,136  & 1,136 \\
						& (353) & (359) & (352) &       & (459) & (472) & (458) &       & (730) & (752) & (728) \\
						\hline
						RA    & --1,301*** & --830** & --1,041*** &       & --230  & 402   & 27    &       & --831  & --264  & --498 \\
						& (350) & (360) & (358) &       & (408) & (426) & (428) &       & (583) & (596) & (591) \\
						\hline
						& \multicolumn{11}{c}{MLE weights} \\
						\cline{2-12}
						IPW   & --1,108*** & --732  & --685  &       & 188   & --34   & 97    &       & --516  & --495  & --337 \\
						& (409) & (534) & (523) &       & (459) & (845) & (793) &       & (611) & (781) & (740) \\
						NIPW  & --1,022** & --564  & --558  &       & 155   & 481   & 502   &       & --515  & --223  & --165 \\
						& (398) & (487) & (485) &       & (452) & (672) & (653) &       & (607) & (718) & (700) \\
						AIPW  & --859** & --613  & --575  &       & 247   & 409   & 584   &       & --434  & --244  & --124 \\
						& (399) & (513) & (504) &       & (449) & (779) & (727) &       & (605) & (753) & (718) \\
						NAIPW & --871** & --626  & --597  &       & 253   & 408   & 514   &       & --434  & --246  & --148 \\
						& (396) & (496) & (491) &       & (451) & (691) & (663) &       & (605) & (724) & (701) \\
						IPWRA & --908** & --590  & --599  &       & 247   & 531   & 533   &       & --443  & --173  & --143 \\
						& (394) & (467) & (470) &       & (452) & (581) & (577) &       & (607) & (682) & (677) \\
						\hline
						& \multicolumn{11}{c}{IPT weights} \\
						\cline{2-12}
						IPW   & --901** & --591  & --599  &       & 253   & 520   & 524   &       & --441  & --176  & --144 \\
						& (394) & (467) & (470) &       & (452) & (588) & (582) &       & (607) & (683) & (677) \\
						NIPW  & --901** & --591  & --599  &       & 253   & 520   & 524   &       & --441  & --176  & --144 \\
						& (394) & (467) & (470) &       & (452) & (588) & (582) &       & (607) & (683) & (677) \\
						AIPW  & --901** & --591  & --599  &       & 253   & 520   & 524   &       & --441  & --176  & --144 \\
						& (394) & (467) & (470) &       & (452) & (588) & (582) &       & (607) & (683) & (677) \\
						NAIPW & --901** & --591  & --599  &       & 253   & 520   & 524   &       & --441  & --176  & --144 \\
						& (394) & (467) & (470) &       & (452) & (588) & (582) &       & (607) & (683) & (677) \\
						IPWRA & --901** & --591  & --599  &       & 253   & 520   & 524   &       & --441  & --176  & --144 \\
						& (394) & (467) & (470) &       & (452) & (588) & (582) &       & (607) & (683) & (677) \\
						\hline
						Linear & \checkmark & & & & \checkmark & & & & \checkmark & & \\
						DW & & \checkmark & & & & \checkmark & & & & \checkmark & \\
						ADW & & & \checkmark & & & & \checkmark & & & & \checkmark \\
						&       &       &       &       &       &       &       &       &       &       &  \\
						Observations & 16,417 & 16,417 & 16,417 & & 16,252 & 16,252 & 16,252 & & 16,134 & 16,134 & 16,134 \\
						\hline
					\end{tabular}
				\end{footnotesize}
				\begin{footnotesize}
					\begin{tablenotes}[flushleft]
						\item \textit{Notes:} The source of data is \cite{LaLonde1986}. The outcome of interest is real earnings in 1978. ``TWFE'' corresponds to the OLS estimation of a panel data specification with outcomes measured in 1975 and 1978, a ``treatment'' indicator, as well as unit and year fixed effects. ``RA'' corresponds to the regression adjustment estimator, which is based on the OLS estimation of a fully interacted linear model with transformed outcomes. The remaining estimators use transformed outcomes and a logit model for the propensity score, and are defined in Section \ref{sec:equivalence}. ``Linear'' corresponds to a specification in which all covariates are included linearly. ``DW'' corresponds to a specification that includes all the covariates in the linear specification as well as an indicator for zero earnings in 1974, age squared, age cubed divided by 1000, years of schooling squared, and an interaction between years of schooling and real earnings in 1974. ``ADW'' corresponds to a specification that includes all the covariates in the DW specification as well as interactions between married and real earnings in 1974 and between married and zero earnings in 1974. Standard errors are in parentheses.
						\item *Statistically significant at the 10\% level; **at the 5\% level; ***at the 1\% level.
					\end{tablenotes}
				\end{footnotesize}
			\end{threeparttable}
		\end{adjustwidth}
	\end{table}
	
	A large number of papers, originating with \cite{LaLonde1986}, combine experimental data from the evaluation of the National Supported Work (NSW) program with a nonexperimental comparison group from the Current Population Survey (CPS) or the Panel Study of Income Dynamics (PSID)\@. The premise of this literature is that a successful nonexperimental estimation method should closely replicate the experimental estimate of the effect of the program when combining the original treatment group with an artificial comparison group \citep[see, e.g.,][]{LaLonde1986,DW1999} or the ``effect'' of zero when combining the latter with the original control group \citep[see, e.g.,][]{ST2005}.
	
	In our application, we closely follow a recent reanalysis of these data in \cite{SAZ2020}, who restrict their attention to samples combining the CPS comparison group and variants of the original control group, previously analyzed by \cite{LaLonde1986}, \cite{DW1999} (the ``DW'' sample), and \cite{ST2005} (the ``early RA'' sample). As is standard in this literature, the outcome of interest is real earnings in 1978. Because \cite{SAZ2020} focus on various difference-in-differences estimators, they often use the transformed outcome, $Y_{i2}-Y_{i1}$; here, this is equal to the difference between real earnings in 1978 and real earnings in 1975. The baseline covariates include age, years of education, real earnings in 1974, and indicator variables for less than 12 years of education, being married, being Black, and being Hispanic. Other covariate specifications also include additional higher-order and interaction terms.
	
	Table \ref{tab:lalonde} replicates every estimate and standard error in \citeauthor{SAZ2020}'s Table 3, while also reporting a number of additional results.\footnote{``TWFE'' corresponds to $\hat{\tau}^{fe}$ in \cite{SAZ2020}. This is the OLS estimate from a panel data specification with real earnings in 1975 and 1978, a ``treatment'' indicator, and unit and year fixed effects. ``RA'' corresponds to $\hat{\tau}^{reg}$ in \cite{SAZ2020}. IPW with MLE weights corresponds to $\hat{\tau}^{ipw,p}$. NIPW with MLE weights corresponds to $\hat{\tau}^{ipw,p}_{std}$. NAIPW with MLE weights corresponds to $\hat{\tau}^{dr,p}$. Finally, all the estimates in the ``IPT weights'' panel are identical to \citeauthor{SAZ2020}'s preferred estimator, $\hat{\tau}^{dr,p}_{imp}$.} The bottom line is that weighting and doubly robust estimators perform quite well in replicating the true effect of zero; except for the simplest covariate specification applied to the \citeauthor{LaLonde1986} sample, none of these estimates are significantly different from the true effect. In addition, \cite{SAZ2020} argue that their preferred estimator based on IPT weights tends to have smaller standard errors than estimators based on MLE weights. While this is true, Table \ref{tab:lalonde} also illustrates our previous point that \citeauthor{SAZ2020}'s (\citeyear{SAZ2020}) estimator might be unnecessarily complex; when using the IPT weights, even the simplest ``unnormalized'' IPW estimator is numerically equivalent to it. Our standard errors, obtained together with the point estimates using our companion Stata package, \texttt{teffects2}, are also identical to those reported by \cite{SAZ2020}.

	\section{Conclusion}
	\label{sec:conclusion}
	
	Applied researchers face many ways to adjust for covariates, but in this paper, we show that several popular estimators are in fact identical under a simple condition. Specifically, our results assume that the propensity score is estimated using inverse probability tilting, a method of moments approach developed by \cite{EGP2008} and \cite{GPE2012,GPE2016}. Estimators based on or equivalent to this approach have already become popular in difference-in-differences settings \citep[cf.][]{SAZ2020} and outside economics \citep[cf.][]{Hainmueller2012,IR2014}. Our results, simplifying the set of alternative estimators available to researchers, offer a novel rationale for adopting this approach in various contexts, such as under unconfoundedness and in instrumental variables and difference-in-differences settings.

	\pagebreak
	\onehalfspacing
	
	\section*{Appendix}
	
	\paragraph{Proof of Proposition \ref{prop:ATE}}
	
	Consider estimating $\mu_{1}$; the argument for $\mu_{0}$ follows in the same way. First, with $\hat{p}_{i} \equiv p ( \mathbf{X}_{i} \mathbf{\hat{\gamma}}_{1,ipt} )$,
	equation \eqref{eq:iptnorm1} implies that
	\begin{equation*}
		N^{-1} \sum_{i=1}^{N} W_{i} / \hat{p}_{i} \; = \; 1.
	\end{equation*}
	The IPT estimator of $\mu_{1}$ is the first term in \eqref{eq:ipt_ate}. Now, consider the AIPT estimator in \eqref{eq:aipt_mu1}. Simple algebra shows it can be expressed as
	\begingroup
	\allowdisplaybreaks
	\begin{eqnarray*}
		\hat{\mu}_{1,aipt} &=& N^{-1} \sum_{i=1}^{N} \frac{W_{i}Y_{i}}{\hat{p}_{i}} \; - \; N^{-1} \sum_{i=1}^{N} \frac{W_{i} \mathbf{X}_{i} \mathbf{\hat{\beta}}_{1}}{\hat{p}_{i}} \; + \; N^{-1} \sum_{i=1}^{N} \mathbf{X}_{i} \mathbf{\hat{\beta}}_{1} \\
		&=& \hat{\mu}_{1,ipt} \; - \; N^{-1} \sum_{i=1}^{N} \frac{W_{i} \mathbf{X}_{i} \mathbf{\hat{\beta}}_{1}}{\hat{p}_{i}} \; + \; \mathbf{\bar{X} \hat{\beta}}_{1} \\
		&=& \hat{\mu}_{1,ipt} \; + \; \left[ \mathbf{\bar{X}} - N^{-1} \sum_{i=1}^{N} \frac{W_{i} \mathbf{X}_{i}}{\hat{p}_{i}} \right] \mathbf{\hat{\beta}}_{1} \\
		&=& \hat{\mu}_{1,ipt},
	\end{eqnarray*}
	\endgroup
	where the last equality uses \eqref{eq:gpemoments_sample} (with an intercept explicitly included).
	
	Now, consider the IPTRA estimator in \eqref{eq:iptra_mu1}. Given that $1 \in \mathbf{X}_{i}$, the first-order condition for the WLS estimator of $\mathbf{\beta}_{1}$, following from \eqref{eq:problem_wls1}, is easily seen to imply that
	\begin{equation*}
		N^{-1} \sum_{i=1}^{N} \frac{W_{i}Y_{i}}{\hat{p}_{i}} \; = \; \left( N^{-1} \sum_{i=1}^{N} \frac{W_{i} \mathbf{X}_{i}}{\hat{p}_{i}} \right) \mathbf{\tilde{\beta}}_{1}.
	\end{equation*}
	The term on the left is, again, $\hat{\mu}_{1,ipt}$. For the term on the right, use the IPT moment conditions in \eqref{eq:gpemoments_sample}, as before:
	\begin{equation*}
		\hat{\mu}_{1,ipt} \; = \; N^{-1} \sum_{i=1}^{N} \frac{W_{i}Y_{i}}{\hat{p}_{i}} \; = \; \left( N^{-1} \sum_{i=1}^{N} \frac{W_{i} \mathbf{X}_{i}}{\hat{p}_{i}} \right) \mathbf{\tilde{\beta}}_{1} \; = \; \mathbf{\bar{X}\tilde{\beta}}_{1} \; = \; \hat{\mu}_{1,iptra}.
	\end{equation*}
	Repeating the same argument for $\mu_{0}$ completes the proof.
	
	\paragraph{Proof of Proposition \ref{prop:ATT}}
	
	Recall that $\mu_{0|1} \equiv \e [ Y(0) | W=1 ]$. Redefine $\hat{p}_{i}$ as $\hat{p}_{i} \equiv p ( \mathbf{X}_{i} \mathbf{\hat{\gamma}}_{0,ipt} )$. Using simple algebra, the AIPW estimator of $\mu_{0|1}$ using IPT weights, given in \eqref{eq:aipwcbps_mu01}, can be written as
	\begin{eqnarray*}
		\hat{\mu}_{0|1,aipt} &=& N_{1}^{-1} \sum_{i=1}^{N} \frac{\hat{p}_{i} \left( 1-W_{i} \right) Y_{i}}{1-\hat{p}_{i}} \; - \; N_{1}^{-1} \sum_{i=1}^{N} \frac{\hat{p}_{i} \left( 1-W_{i} \right) \mathbf{X}_{i} \mathbf{\hat{\beta}}_{0}}{1 - \hat{p}_{i}} \; + \; \mathbf{\bar{X}}_{1} \mathbf{\hat{\beta}}_{0} \\
		&=& \hat{\mu}_{0|1,ipt} \; + \; \left[ \mathbf{\bar{X}}_{1} - N_{1}^{-1} \sum_{i=1}^{N} \frac{\hat{p}_{i} \left( 1-W_{i} \right) \mathbf{X}_{i}}{1 - \hat{p}_{i}} \right] \mathbf{\hat{\beta}}_{0} \\
		&=& \hat{\mu}_{0|1,ipt},
	\end{eqnarray*}
	where the final equality follows from equation \eqref{eq:irmoments_att_sample}, the IPT moment conditions for estimating the ATT\@.
	
	For the IPWRA estimator using the IPT weights, note that the first-order condition for $\mathbf{\tilde{\beta}}_{0}$, following from \eqref{eq:problem_wls0}, is
	\begin{equation*}
		\sum_{i=1}^{N} \frac{\hat{p}_{i} \left( 1-W_{i} \right)}{1 - \hat{p}_{i}} \mathbf{X}_{i}^{\prime} \left( Y_{i} - \mathbf{X}_{i} \mathbf{\tilde{\beta}}_{0} \right) \; = \; \mathbf{0}.
	\end{equation*}
	Focusing on the first element $1 \in \mathbf{X}_{i}$ and dividing by $N_{1}$, we can write
	\begin{equation*}
		N_{1}^{-1} \sum_{i=1}^{N} \frac{\hat{p}_{i} \left( 1-W_{i} \right) Y_{i}}{1-\hat{p}_{i}} \; = \; N_{1}^{-1} \sum_{i=1}^{N} \frac{\hat{p}_{i} \left( 1-W_{i} \right) \mathbf{X}_{i} \mathbf{\tilde{\beta}}_{0}}{1 - \hat{p}_{i}}
	\end{equation*}
	or, again using \eqref{eq:irmoments_att_sample},
	\begin{equation*}
		\hat{\mu}_{0|1,ipt} \; = \; \mathbf{\bar{X}}_{1} \mathbf{\tilde{\beta}}_{0} \; = \; \hat{\mu}_{0|1,iptra}.
	\end{equation*}
	This completes the proof.

	\pagebreak
	\singlespacing
	
	\setlength\bibsep{0pt}
	\bibliographystyle{econometrica}
	\bibliography{SUW_references}
	
\end{document}